\documentclass[useAMS]{mn2e}

\usepackage{graphicx}
\title[Accretion in The Necklace]{A carbon dwarf wearing a Necklace: first proof of accretion in a post-common-envelope binary central star of a planetary nebula with jets\thanks{Based on observations made with the Very Large Telescope (VLT) at Paranal Observatory under programme 085.D-0629(A) and service observations made with the 4.2 m William Herschel Telescope (WHT) operated on the island of La Palma by the Isaac Newton Group in the Spanish Observatorio del Roque de los Muchachos of the Instituto de Astrof\'\i sica de Canarias.}}

\author[B. Miszalski et al.]{Brent Miszalski,$^{1,2}$\thanks{E-mail: brent@saao.ac.za} Henri M.~J. Boffin$^{3}$ and Romano L. M. Corradi$^{4,5}$\\
$^{1}$South African Astronomical Observatory, PO Box 9, Observatory, 7935, South Africa\\
$^{2}$Southern African Large Telescope Foundation, PO Box 9, Observatory, 7935, South Africa\\
$^{3}$European Southern Observatory, Alonso de Cordova 3107, Casilla 19001, Santiago, Chile\\
$^{4}$Instituto de Astrof{\'{\i}}sica de Canarias, E-38200 La Laguna, Tenerife, Spain\\
$^{5}$Departamento de Astrof{\'{\i}}sica, Universidad de La Laguna, E-38206 La Laguna, Tenerife, Spain\\
}
\begin{document}

\date{Accepted Received ; in original form }

\pagerange{\pageref{firstpage}--\pageref{lastpage}} \pubyear{2012}

\maketitle

\label{firstpage}
\begin{abstract}
   The formation of collimated outflows or jets in planetary nebulae (PNe) is not well understood. There is no evidence for active accretion disks in PNe making it difficult to decide which of several proposed jet formation scenarios may be correct. A handful of wide binary central stars of PNe are known to have accreted carbon and slow neutron capture (s-process) enhanced material, the immediate progenitors of barium stars, however no close binary analogues are known to have passed through a common-envelope (CE) phase. Here we present spectroscopy of The Necklace taken near lightcurve minimum that for the first time reveals a carbon-rich (C/O$>$1) companion, a carbon dwarf, in a post-CE central star. As unevolved stars do not produce carbon, the chemical enhancement of the secondary can only be explained by accretion from the primary. Accretion most likely happened prior to the CE phase via wind accretion as not enough material can be accreted during the short CE phase. The pair of jets in The Necklace, which are observed to be older than the PN, are therefore likely to have been launched from an accretion disk around the companion during this early accretion phase. This discovery adds significant weight to the emerging scenario that jets in post-CE PNe are primarily launched by an accretion disk around a main-sequence companion before the CE phase.  
\end{abstract}


   \section{Introduction}
   \label{sec:intro}
   A small group of planetary nebulae (PNe) are known to harbour binary central stars where a sub-giant or giant companion is enriched in carbon and slow neutron capture process (s-process) elements (Bond, Ciardullo \& Meakes 1993; Th\'evenin \& Jasniewicz 1997; Bond, Pollacco \& Webbink 2003; Miszalski et al. 2012). These are the progenitors of the barium stars (Bidelman \& Keenan 1951), caught during the short-lived phase ($\sim$$10^4$ yr) when the PN ejected by the white dwarf (WD) is still visible. The wind of the WD precursor, loaded with carbon and s-process rich material dredged up via thermal pulses during its asymptotic giant branch (AGB) phase, pollutes the companion which accretes the enriched material (Boffin \& Jorissen 1988). The wind accretion probably occurs while the companion is on the main sequence, as shown by the existence of several barium dwarfs (Gray et al. 2011 and ref. therein).

   The existence of barium central stars constitutes the only firm evidence for mass transfer and accretion onto non-WD companions in PNe. Observations of close binary central stars, i.e. those that have passed through a common-envelope (CE) phase and have orbital periods less than $\sim$1 day (e.g. De Marco et al. 2008; Miszalski et al. 2009a), show no evidence for rapid variability (flickering) or spectroscopic features that could be attributed to accretion (see e.g. Exter et al. 2005). The presence of collimated outflows or jets surrounding several systems (Mitchell et al. 2007; Miszalski et al. 2009b, 2011a, 2011b; Corradi et al. 2011; Boffin et al. 2012) therefore appears to challenge the suspected strong connection between accretion disks and jets (Morris 1987; Soker \& Livio 1994). These jets were launched either from an accretion disk that is no longer present or alternatively a dynamo effect (Nordhaus \& Blackman 2006; Nordhaus, Blackman \& Frank 2007). An accretion disk may have formed prior to the CE phase via wind accretion from the AGB primary (Boffin, Theuns \& Jorissen 1994; Theuns, Boffin \& Jorissen 1996; Mastrodemos \& Morris 1998; Soker \& Rappaport 2000; de Val-Borro, Karovska \& Sasselov 2009; Huarte-Espinosa \& Frank 2012), during the start of the CE infall phase (Soker 2004) or perhaps even after the CE phase (Soker \& Livio 1994). 

   A clearly defined sample of PNe with jets and binary central stars is required to decide between which of the possible jet formation scenarios are operating in PNe. Only recently has this task become possible with the discovery of several post-CE nebulae with jets. Previous investigations have relied upon pre-PNe which are a heterogeneous group with scant information on their binary status making them difficult to place amongst the binary-dependent jet scenarios (e.g. Huggins 2007). None are post-CE systems and there is no support for a large fraction of binaries in pre-PNe (Hrivnak et al. 2011). Several insights into jet formation scenarios have already emerged from the newly forged post-CE sample. Longslit observations of some jets around post-CE nebulae indicate the jets were probably ejected before their main nebulae (Mitchell et al. 2007; Miszalski et al. 2011a; Corradi et al. 2011; Boffin et al. 2012). Another fundamental clue comes from point-symmetric outflows of PNe. Simulations can recreate these complex outflows with a precessing accretion disk around the secondary launching jets (e.g. Cliffe et al. 1995; Raga et al. 2009). While there are multiple examples of such PNe, none were known to have a binary nucleus until the landmark discovery of a post-CE binary nucleus in the archetype of this class Fleming~1 (Boffin et al. 2012). The nebular and central star characteristics of Fleming~1 together form a strong validation of this model.
   
   If accretion disks were responsible for launching the jets of post-CE PNe, then the aforementioned observations suggest they were temporary. To further test our hypothesis we require the detection of some vestigial signature of accretion in a post-CE binary central star. 
   The clearest proof for accretion would be a polluted main-sequence companion with an atmosphere strongly enriched by accreted material. The secondary is best observed during the primary eclipse of eclipsing systems, when the secondary eclipses the primary and its irradiated zone is obscured, but it can also be observed at light minimum in non-eclipsing irradiated systems if the luminosity of the primary is low enough not to overpower the secondary. 

Unfortunately, the few direct observations of other post-CE secondaries in PNe have not revealed any such enrichment (e.g. Ferguson et al. 1987; Walton, Walsh \& Pottasch 1993). More recently, Wawrzyn et al. (2009) modelled the irradiated C and N emission lines in A~63 and found them to be 3--5 times too strong, giving the explanation that they were possibly enhanced by the accretion of C- and N-rich material from a polluted AGB wind. Wawrzyn et al. (2009) remarked that further work is required to independently confirm this model-dependent result and that this may involve analysis of the non-irradiated side of the secondary. Remaining evidence for accretion onto post-CE secondaries is more indirect. It is suspected to be responsible for the frequently observed larger radii of secondaries in post-CE central stars that are $\ga$2 times larger than their main-sequence equivalents (e.g. Af{\c s}ar \& Ibano{\v g}lu 2008 and ref. therein). Another consequence is thought to be the induction of rapid rotation in secondaries as supported by the detection of X-ray emitting coronae in two post-CE central stars and one barium central star (Montez et al. 2010). 

   In this \emph{Letter} we report on the detection of a carbon dwarf secondary in a post-CE central star binary as the first firm proof for a previous accretion phase. Section \ref{sec:obs} describes the observations of the central star of The Necklace (PN G054.2$-$03.4; Viironen et al. 2009; Corradi et al. 2011) which is a post-CE binary with an orbital period of 1.16 d and an extremely large irradiation effect with an $I$-band amplitude of 0.75 mag. Section \ref{sec:disc} describes the carbon dwarf nature of the secondary, the implications of accretion for the formation of the observed jets in The Necklace, and we conclude in Section \ref{sec:conclusion}. 
\vspace{-0.7cm}
   \section{observations}
   \label{sec:obs}
   The central star of The Necklace was observed with the FORS2 instrument (Appenzeller et al. 1998) as part of the VLT visitor mode program 085.D-0629(A) on 18 June 2010. The blue optimised E2V detector was used in combination with the 1200g grism to give wavelength coverage from 4182--5664 \AA. One 40 minute spectrum was obtained at an airmass of 1.34 with the 0.5\arcsec\ slit. During the exposure the orbital phase varied between $\phi_\mathrm{VLT}=$ 0.057--0.081 according to the ephemeris of Corradi et al. (2011), very close to the lightcurve minimum (i.e. near to inferior conjunction of the cool companion). The detector was readout with $1\times1$ binning to give a dispersion of 0.72 \AA\ pixel$^{-1}$ and a resolution measured from arc lines of 1.2 \AA\ (full-width at half-maximum, FWHM) near 4900 \AA. The data were reduced using the ESO FORS pipeline and one-dimensional spectra were extracted with the \textsc{iraf} task \textsc{apall}. A signal-to-noise ratio (S/N) of 16 was achieved at $\lambda$4800 \AA. Flux calibration was performed in the usual fashion with a 120 s exposure of the CSPN of NGC~7293 (Oke 1990) observed with the same setup on 16 June 2010. 

The central star was reobserved in service time on 27 March 2012 with ACAM (Benn, Dee \& Ag{\'o}cs 2008) on the 4.2-m WHT telescope. A 1.0\arcsec\ longslit was aligned with the parallactic angle and the 400 lines mm$^{-1}$ volume-phase holographic grating was used with the GG495 order-blocking filter
to avoid second-order contamination at the red wavelengths considered. 
Two exposures of 30 min each were taken and added together during the reduction stage, performed with standard \textsc{iraf} routines. The observations were made when the orbital phase varied between $\phi_\mathrm{WHT}=$ 0.09--0.13, again close to the lightcurve minimum. The spectrophotometric standard star BD$+$33$^\circ$2642 from Oke (1990) was observed during the same night and used for the flux calibration. 
The reciprocal dispersion was 3.4~\AA~pix$^{-1}$ and the resolution $\sim$13~\AA\ (FWHM). A S/N of 40 was achieved at $\lambda$6400~\AA. Slit losses between the VLT and WHT spectra mean the flux scales are not comparable.
   
\begin{figure*}
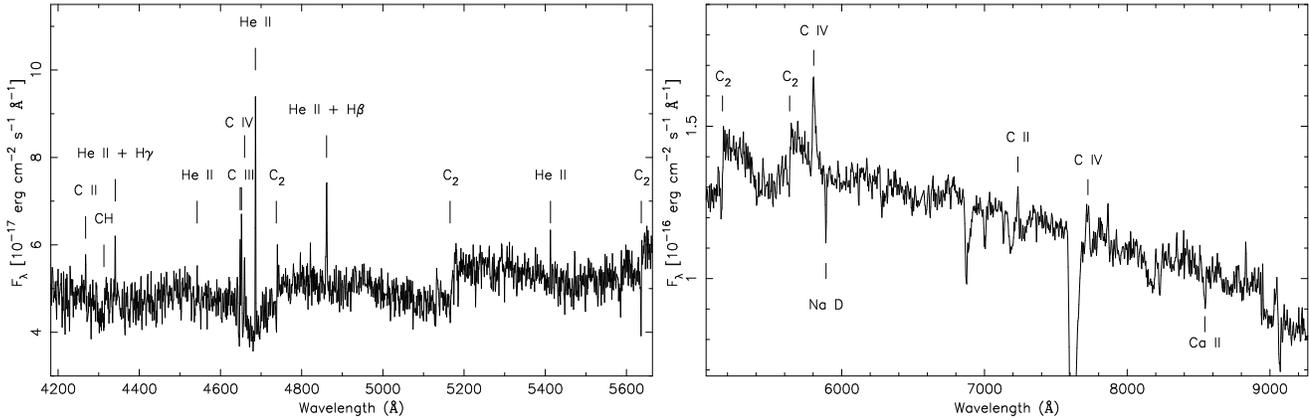

      \begin{center}
         \includegraphics[scale=0.375,angle=270]{necklace.ps}
         \includegraphics[scale=0.375,angle=270]{ACAM.ps}
      \end{center}
      \caption{VLT FORS2 (left) and WHT ACAM (right) spectra of The Necklace central star obtained near light minimum. The nebula emission lines of [O~III] and H$\alpha$ were interpolated over. The C$_2$ bands prove the secondary is carbon-rich (C/O $>$ 1).}
      \label{fig:spec}
   \end{figure*}

Figure \ref{fig:spec} shows both nebula-subtracted, flux-calibrated spectra with prominent features identified that belong to either the secondary or its irradiated atmosphere. These include the C$_2$ Swan bands at $\lambda\lambda$4737, 5165 and 5636 \AA, and the CH band at $\lambda$4312, while the irradiation zone emission lines are from C~II ($\lambda\lambda$4267 and 7235 \AA), C~III ($\lambda\lambda$4647 and 4650 \AA), C~IV ($\lambda\lambda$4658, 5801 and 5812, and 7726 \AA) and He~II ($\lambda\lambda$4338, 4540, 4686, 4860 and 5412 \AA). Stellar absorption lines of Na~I and Ca~II are also present. No prominent features are seen from CN or s-process elements (e.g. Ba~II $\lambda$4554). The blue continuum of the ACAM spectrum is due to the dominance of the pre-WD primary and we estimate that each component contributes approximately equally to the combined spectrum. 

\vspace{-0.7cm}
   \section{Discussion}
   \label{sec:disc}
   \subsection{A carbon dwarf polluted via accretion}
   We attribute the strong C$_2$ bands visible in Fig. \ref{fig:spec} to a main-sequence secondary enriched in carbon. The C$_2$ bands cannot belong to a DQ WD primary (e.g. Dufour, Bergeron \& Fontaine 2005) because at the $4.6\pm1.1$ kpc distance of The Necklace (Corradi et al. 2011) the primary would be exceedingly faint at the typical luminosities of DQ WDs ($M_V\ga13$ mag), the temperature would be too cool to ionise the nebula and the surface gravity too high. A DQ secondary is ruled out by the presence of Na~I and Ca~II absorption lines from a late-type star. Since no carbon is produced by main-sequence hydrogen burning, this material must have been accreted from a carbon-rich wind of the primary when it was an AGB star. 
   
   The secondary star most resembles the carbon dwarfs (Green \& Kurtz 1998; Sect. 3 of Wallerstein \& Knapp 1998) that were first discovered by Dahn et al. (1977). It is the first documented case in a short-period post-CE binary (Wallerstein \& Knapp 1998). As in Barium stars, carbon dwarfs are enriched via accretion of carbon and s-process rich winds from their AGB companions that have since evolved into WDs (e.g. Heber et al. 1993; Liebert et al. 1994; Green \& Margon 1994). The presence of C$_2$ bands requires an atmospheric composition with C/O $>$ 1 by definition (Wallerstein \& Knapp 1998). Parallax studies find the secondaries to be very faint with $M_V$ = 9.5--10.0 mag (Dearborn et al. 1986; Harris et al. 1998). Most belong to the Galactic Halo, although some disk members are known (e.g. Heber et al. 1993), and they are probably the most numerous of mass transfer binaries (Green et al. 1992).

   The binary population synthesis study of de Kool \& Green (1995) reproduced the observed space density of carbon stars reasonably well, notwithstanding the uncertainties associated with the choice of model initial conditions. Both the Halo and Disk populations are modelled. The radial velocity, nebular chemical abundances and small 270 pc height below the Galactic Plane of The Necklace (Corradi et al. 2011) all point to a Galactic Disk membership. Given the strong selection effects involved with comparing one object against the de Kool \& Green (1995) models, it is difficult to gauge how well The Necklace may fit the observed population of carbon dwarfs. The 1.16 d orbital period is only just compatible with the shorter component of the bimodal orbital period distribution of de Kool \& Green (1995) that corresponds to post-CE carbon dwarfs. 
   
   We may also anticipate the secondary to share the low luminosity of observed carbon dwarfs. A weighted average of $M_V=9.72\pm0.16$ mag from three carbon dwarfs (Harris et al. 1998) is however not consistent with observations of The Necklace. If we assume the carbon dwarf contributes a fraction $f=0.5$ of the total intensity and an average $(V-I)_0=1.51$ mag colour for carbon dwarfs (Harris et al. 1998), then at lightcurve minimum [$I_0(\mathrm{min}) = 17.39$ mag and $A(I)=0.56$ mag (Corradi et al. 2011)] an unlikely distance of $\sim$1 kpc would be required. On the other hand, the post-CE nature of The Necklace means that we may expect the secondary to have a radius up to 2.5 times greater than a normal carbon dwarf (Af{\c s}ar \& Ibano{\v g}lu 2008). This means the luminosity could be 2.0 mag brighter than a typical carbon dwarf. It could be even more luminous if the secondary were also more massive than a carbon dwarf and this might be the case given the relatively massive secondary of K~1-2 which demonstrates a similarly high-amplitude irradiation effect to the Necklace (Exter et al. 2003; De Marco et al. 2008). Given the unknown values of the secondary radius and luminosity, and to a lesser extent $f$, it is not possible to derive a meaningful and consistent distance estimate. It does however seem that the secondary is more likely to be an overheated, enlarged carbon dwarf at the estimated distance of $4.6\pm1.1$ kpc (Corradi et al. 2011), rather than a typical faint carbon dwarf at $\sim$1 kpc.

\vspace{-0.6cm}
   \subsection{Wind accretion and jet formation}
   \label{sec:when}
   The carbon dwarf secondary is the first clear-cut proof of accretion onto a main-sequence secondary in a post-CE binary central star. If an accretion disk formed when this accretion occurred, it would have likely launched the jets in The Necklace, given the strong connection between astrophysical jets and accretion disks (e.g. Pudritz et al. 2007). Knowing when the accretion phase occurs is therefore crucial to decide between the various possible jet formation scenarios for PNe proposed in the literature. We consider three main scenarios to produce an accretion disk required to launch the observed jets:
   
   \begin{enumerate}
      \item Wind accretion prior to the CE phase resulting in an accretion disk around the secondary (see references listed in Sect. \ref{sec:intro}),
      \item Short-lived accretion when the secondary orbits just inside the AGB star's envelope (e.g. Soker 2004),
      \item After the CE phase the out of thermal equilibrium secondary transfers matter accreted during the CE phase back to the pre-WD primary forming an accretion disk around the primary. Another possibility may be reaccretion of a partially ejected CE, although it is unclear where such material may settle. Both scenarios were discussed by Soker \& Livio (1994).
   \end{enumerate}

   The strongest evidence against scenarios (ii) and (iii) is that several jet systems of post-CE PNe are known to be several 10$^3$ yrs older than the inner nebulae (Mitchell et al. 2007; Miszalski et al. 2011a; Corradi et al. 2011; Boffin et al. 2012). This time difference in The Necklace corresponds to $\sim$5000 yrs at $d=4.6$ kpc. If these scenarios were correct, then the jets should be coeval with the ejected CE, launched during the extremely short window of the CE interaction ($\sim$1--10 yr, Passy et al. 2012). Even if one were to account for the distance dependence of these timescales, and possibly nebula evolutionary effects (Huggins 2007), it is difficult to resolve the two orders of magnitude timescale difference. We discard the Nordhaus \& Blackman (2006) dynamo in this case because (a) it occurs during the CE infall phase that is inconsistent with the observed timescale difference, and (b) the non-accretion driven nature of the dynamo may not produce a carbon dwarf secondary. 
   
Scenario (ii) is also particularly sensitive to the amount of accretion during the CE interaction. A minimum amount of $\Delta M_2$ must be accreted during this brief stage to change the assumed solar composition $(C/O)_i = 1/3$ to the observed value $(C/O)_f > 1$. The latter is the outcome of accreted carbon-rich material being diluted into the secondary's envelope of mass $M_{2,e} = 0.35 \left[ \left(1.25-M_2\right)/0.9\right]^2$ (Hurley, Pols \& Tout 2000). The final composition is then 

\begin{equation}
   \left(\frac{C}{O}\right)_f = \frac{\left(C/O\right)_i + \eta \left(C/O\right)_\mathrm{AGB} \left(\Delta M_2/M_{2,e}\right)}{1+ \eta \left(\Delta M_2/M_{2,e}\right)} \label{eqn}
   \end{equation}

   where $\left(C/O\right)_\mathrm{AGB}$ is the ratio in the AGB star which has a metallicity $\eta=\left(O_\mathrm{AGB}/O_\mathrm{sun}\right)$. Assuming a typical $(C/O)_\mathrm{AGB}$ = 1.5--3.0 for AGB stars at the end of their life (Lambert et al. 1986) and $\eta=1$ (solar metallicity), we find from Eqn. \ref{eqn} that we need $\left(\Delta M_2/M_{2,e}\right) \sim 1/3$, i.e. $ \Delta M_2 \sim 0.48 (1.25 - M_2 )^2$. Thus $\Delta M_2$=0.03--0.35 for $M_2=$1.0--0.4 $M_\odot$, respectively. The most advanced simulations of the in-spiral part of the CE phase by Ricker \& Taam (2008, 2012) and Passy et al. (2012) predict a negligible $\Delta M_2\sim$10$^{-3}$ $M_\odot$, well below our estimated $\Delta M_2$. Launching jets during this phase may also be very difficult unless it happens very early in the interaction (Soker 2004). 

   The carbon-rich composition of a post-CE binary is challenging to explain. De Marco (2009) suggested that the CE interaction would make it statistically difficult for such a binary to form, in which the companion terminates the evolution of the AGB star before it becomes a carbon star. There are however plausible circumstances to produce The Necklace. Solar metallicity stellar evolution models of Marigo et al. (in prep.) show a significant increase in the AGB star radius when the C/O ratio exceeds unity. The radius also increases after each thermal pulse, a necessary ingredient to trigger a CE interaction (e.g. De Marco et al. 2011). This implies that only some systems will experience a CE phase when the AGB star is carbon-rich. Marigo and colleagues show that for a representative initial mass of 2.6 $M_\odot$ the star spends $\sim$10$^6$ years as a carbon star, achieving a radius of at least 800 $R_\odot$, significantly larger than the maximum size reached during thermal pulses in the oxygen-rich phase ($\le$400 $R_\odot$). In this time the star would lose 0.1--1.0 M$_\odot$ with a 10$^{-6}$--10$^{-7}$ M$_\odot$ yr$^{-1}$ wind, of which the companion would gain 1--30\% in a typical wind accretion scenario, easily matching our estimated $\Delta M_2$. Eventually, the primary would grow to a size that a CE phase is initiated, terminating the AGB evolution of the primary and wind accretion onto the secondary. With these parameters, the initial separation would be 400--1000 $R_\odot$, implying orbital periods 500--2000 days. These periods are typical of symbiotic stars, where substantial wind accretion onto companions is known to occur. A small CE efficiency would then produce a dramatic decrease in the orbital period to the current value.
      
   An alternative way to produce a large amount of accretable carbon might be for the pre-WD to experience a late thermal pulse (LTP; e.g. Sch\"onberner 1979). However, if the companion were to accrete carbon in this scenario and launch jets, then they would have to be younger than the PN. This is not observed in The Necklace. In addition, the radius increase of the primary following the LTP would likely trigger another CE phase before any dredge-up would take place to mix carbon to the surface of the primary, as surface enrichment only occurs when the star has reached the minimum effective temperature and is in the reheating phase (see e.g. Bl\"ocker 2001). This further complicates the possibility of transferring carbon-rich material to the secondary (see also previous discussion of scenarios ii and iii). Finally, there is no visible sign of H-deficient circumbinary material with peculiar abundances associated with an LTP (e.g. Wesson et al. 2008).

   In summary, the wind accretion scenario provides a much simpler and viable solution. Other additional interactions may also occur but scenario (i) would be the main mechanism behind polar outflows in PNe with companions that survive the CE phase.

\vspace{-0.7cm}
\section{Conclusions}
\label{sec:conclusion} 
We have discovered the first example of a mass transfer binary in a post-CE binary central star of a PN. The carbon dwarf secondary was identified in spectra taken near the lightcurve minimum from strong C$_2$ bands. These features require an atmospheric composition with C/O $>$ 1 which can only be caused by accretion of a carbon-rich wind from the AGB-precursor of the observed pre-WD primary. Advanced simulations of the in-spiral part of the CE phase (Ricker \& Taam 2008, 2012; Passy et al. 2012) predict an insufficient amount of accretion to take place during this short time to produce the carbon dwarf composition. The accretion is therefore most likely to have occurred before the CE phase via wind accretion, a process that simulations predict to form an accretion disk around the companion (e.g. Morris 1987). Since the jets of The Necklace are observed to be older than the main PN (Corradi et al. 2011), they were probably launched from such a disk. We therefore predict that the jets may be enhanced in carbon and UV spectroscopy with the \emph{HST} is uniquely placed to measure any overabundance they may have.

\vspace{-0.7cm}
\section*{Acknowledgments}
We thank P. Marigo and colleagues for sharing with us their latest stellar evolution model results in advance of publication. BM thanks Ralf Napiwotzki for discussions about DQ WDs in the very early phases of this project. HMJB wishes to warmly thank Yuri Beletsky and Leonel Rivas for their dedicated support at the VLT during these observations.

\label{lastpage}

\end{document}